\documentstyle[amssymb,aps,prl,floats,twocolumn,epsf]{revtex}

\begin{document}

\wideabs{
\title{Metal-Insulator oscillations in a Two-dimensional Electron-Hole system}
\author{R.J. Nicholas, K. Takashina, M. Lakrimi, B. Kardynal, S. Khym, N.J. Mason}
\address{Department of Physics, Oxford University, \\Clarendon Laboratory, Parks Rd.,
Oxford, OX1 3PU, U.K.}
\author{D.M. Symons, D. K. Maude and J.C. Portal}
\address{Grenoble High Magnetic Field Laboratory, Max Planck Institut f\"{u}r\\
Festk\"{o}rperforschung and Centre National de la Recherche Scientifique\\
 BP166, 38042 Grenoble, Cedex 9, France.}
\date{\today }
\maketitle

\begin{abstract}
The electrical transport properties of a bipolar InAs/GaSb system have been
studied in magnetic field. The
resistivity oscillates between insulating and metallic behaviour while the
quantum Hall effect shows a digital character oscillating from 0 to 1
conducatance quantum e$^{2}$/h. The insulating behaviour is
attributed to the formation of a total energy gap in the system. A novel
looped edge state picture is proposed associated with the appearance of 
a voltage between Hall probes which is symmetric on magnetic field reversal.
\end{abstract}

\pacs{73.40.Hm, 73.61.Ey, 71.30.+h}
}

The quantum Hall effect has become well known since its first observation by
von Klitzing et al\cite{Klitzing} but still raises a number of fundamental
questions, including the role of edge states which ensure the possibility of
dissipationless conduction\cite{buttiker}\cite{chlovskii}. In this letter we
examine the quantum Hall effect and magnetotransport properties of a bipolar
system of coupled electrons and holes and demonstrate that such a system
shows qualitatively different behaviour to that observed for single
carriers. Such systems have generated considerable interest due to the
possibilities of gap formation by both excitonic\cite{Naveh}\cite{Littlewood}
and single particle\cite{deLeon} interactions. The electron-hole system
introduces the new possibility that the edge states of the electron and hole
systems may interact, breaking the normal quantum Hall conditions. The first
observation of quantum Hall plateaux in an electron-hole system by Mendez et
al \cite{Mendez} found that conventional plateaux were formed at quantum numbers
corresponding to the difference in the occupancies of the electron and hole
Landau levels. Subsequently Daly et al\cite{Daly} found that in
superlattices with closely matched electron and hole densities the special
case of zero Hall resistance could be observed. In this letter we examine
the behaviour of insulating states formed in a structure containing one
layer each of electrons and holes which interact via interband mixing. If
mixing occurs between edge states a total energy gap may appear for the
system leading to the insulating behaviour. Several gaps can result from the
mixing between different electron and hole Landau levels and as a result the
system displays oscillatory metallic and insulating behaviour as a function
of magnetic field and the Hall conductivity follows a binary sequence
oscillating from 0 - 1 - 0 conductance quanta.

The samples studied consisted of a single layer of InAs sandwiched between
thick layers of GaSb. This system exhibits a broken gap lineup with the
conduction band edge of the InAs 150meV below the valence band edge of the
GaSb. The samples are grown by Metal Organic Vapour Phase Epitaxy (MOVPE)
and have a relatively low level of impurities so that the majority of charge
carriers are created by intrinsic charge transfer from the GaSb layers to
the InAs layer. Typical structures are grown onto semi-insulating GaAs
followed by 2$\mu $m of GaSb to achieve lattice relaxation. The active layer
of InAs is typically 30nm thick, followed by a 90nm GaSb cap. Fitting the
low field magnetoresistance and Hall effect to classical two carrier
formulae give carrier densities of the electrons and holes of order 6.5$%
\times $10$^{15}$ m$^{-2}$ and 4.5$\times $10$^{15}$ m$^{-2}$ respectively.
Evidence from magnetotransport suggests that the structure is asymmetric due
to pinning of the Fermi level $E_{F}$ at the surface so that all of the
holes are on one side of the structure giving Fermi surfaces for the
electrons and holes of similar magnitudes\cite{symons}\cite{Lakrimi}.
Determination of the carrier mobilities is less straightforward due to the
minigap caused by the electron-hole interaction. However, by applying a
large parallel magnetic field, which is known to decouple the bands,\cite
{Lakrimi}\cite{Yang} the mobilities for the decoupled electrons and holes
are estimated to be $\thicksim $ 20 and 1 m$^{2}$V$^{-1}$s$^{-1}$.

Magnetotransport measurements were made using A.C. techniques with a
dilution refrigerator and 15T superconducting magnet, and a $^{3}$He
cryostat and 50T pulsed field magnet system on wide Hall bars of width 0.5
mm. Measurements were made with the magnetic field in both forward ($B+$)
and reverse ($B-$) directions so that the resistances can be separated into
symmetric (S) and antisymmetric (A) parts with respect to field reversal:

$R_{xx}^{S}=(R_{xx}(B+)+R_{xx}(B-))/2=\rho _{xx}\times L/W$

$R_{xx}^{A}=(R_{xx}(B+)-R_{xx}(B-))/2$

$R_{xy}^{S}=(R_{xy}(B+)+R_{xy}(B-))/2$

$R_{xy}^{A}=(R_{xy}(B+)-R_{xy}(B-))/2=\rho _{xy}$

The symmetric part of $R_{xx}$ was used with the length to width ratio $%
(L/W) $ of the bar to give the diagonal resistivity $\rho _{xx}$ and the
antisymmetric part of $R_{xy}$ is taken to be the Hall resistivity, $\rho
_{xy}$ according to the Onsager relations. The antisymmetric part of $R_{xx}$
is generally found to be negligibly small, and the symmetric part of $R_{xy}$
is usually attributed to the (small) admixing of the diagonal resistivity\cite{Dunf}.

\begin{figure}[tbp]
\epsfxsize=3.3in
\epsffile{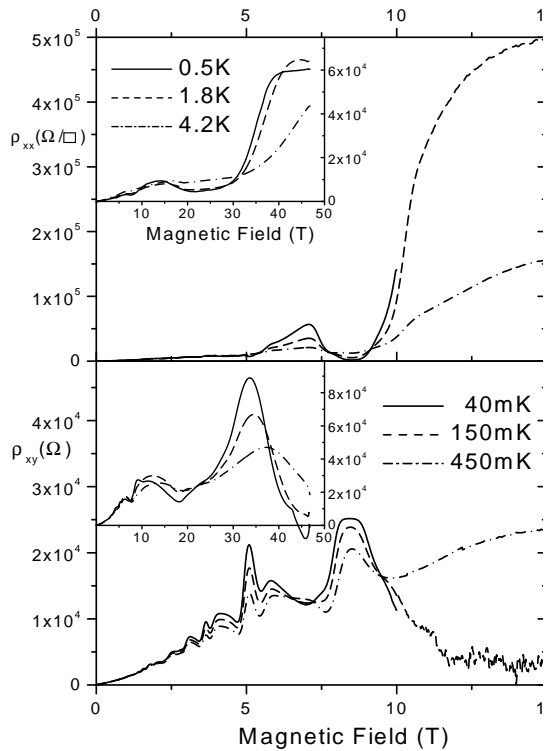}
\caption{The magnetoresistivity components $\protect\rho _{xx}$ and $\protect%
\rho _{xy}$ measured in steady fields (main figure) and pulsed fields
(insets) for temperatures from 40mK to 4.2 K.}
\end{figure}

Fig. 1 shows a typical series of plots of the magneto- and Hall resistivity
at temperatures down to 40mK using steady fields and 500mK in pulsed fields.
The traces show a number of striking features characteristic of two carrier
structures\cite{Nicholas}\cite{Dalton} and are very different from single
carrier systems. There is a very large positive magnetoresistance with
relatively weak oscillatory features at low field. At higher fields there
are strong minima at 8.5T and 22T and large maxima at 7T, 15T and 45T which
increase rapidly with falling temperature. The peak at 15T has a resistivity
which increases to several thousand times its zero field value at 150mK. By
contrast the Hall resistivity shows stronger oscillations at low fields, but
has only one feature which can be attributed to the formation of a quantum
Hall plateau at 8.5T which only appears at the lowest temperatures. Minima
occur where $\rho _{xy}$ approaches zero at 15T by 150mK and 45T by 500mK.

\begin{figure}[tbp]
\epsfxsize=3.3in
\epsffile{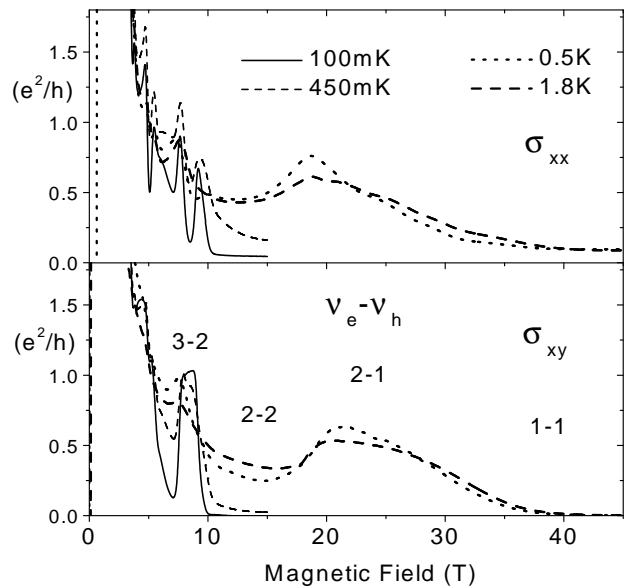}
\caption[figure 2 ]{The conductivity components $\protect\sigma _{xx}$ and 
$\protect\sigma _{xy}$
calculated from the data in Fig. 1. The figures indicate the expected
electron ($\protect\nu _{e}$) and hole ($\protect\nu _{h}$) Landau level
occupancies.}
\end{figure}

The origins of this behaviour are clearer when the resistivity components
are converted to conductivity\cite{Nich2}. Fig. 2 shows that $\sigma _{xy}$
has a low temperature zero above 40T and in the region 10 - 15T,
and approaches zero around 7T. Between these fields one
conductance quantum of $e^{2}/h$ is observed at the lowest temperatures in
the steady field data, and a peak moving towards this value with
falling temperature is observed at around 20T in the pulsed field data. Minima 
occur in $\sigma _{xx}$ when $\rho _{xx}$ has both minima and strong maxima where 
$\rho _{xx}\gg \rho _{xy}$.

An important feature of broken gap systems is that the electron Landau
levels start at a lower energy than the hole levels, and as a function of
magnetic field the two sets of levels must cross. Due to the
finite interband mixing between the valence and conduction bands the levels
anticross leading to the formation of a miniband gap\cite{Lakrimi}\cite{Yang}
\cite{Altarelli} \cite{Chiang}\cite{Cooper}\cite{Vasilev}\cite{Petch}, shown
schematically in fig. 3i. The movement of the levels through each other
leads to oscillatory carrier densities due to the varying density of states,
but the position of the Fermi level varies much less than in single carrier
systems due to charge transfer between the bands. For structures with a
small net doping (i.e. $n_{e}-n_{h}$ is small and constant) the Fermi level
can lie close to the centre of either electron or hole Landau levels or
within the localised tail states of either or both carrier types. In the
simplified schematic fig. 3i the Fermi level is shown as constant. The net
occupancy of the levels will oscillate between zero around positions a
(where $\nu _{e}$ - $\nu _{h}$ = 0 - 0) and c (where $\nu _{e}$ - $\nu _{h}$
= 1 - 1) and 1 for b (where $\nu _{e}$ - $\nu _{h}$ = 1 - 0) and d (where $%
\nu _{e}$ - $\nu _{h}$ = 2 - 1). The particularly unusual features of the
conduction process occur when the Fermi level lies between equal numbers of
electron and hole Landau levels. In this case the system is able to show a
complete energy gap which leads to the appearance of oscillatory insulating
behaviour as a function of magnetic field.

\begin{figure}[tb]
\epsfxsize=3.3in
\epsffile{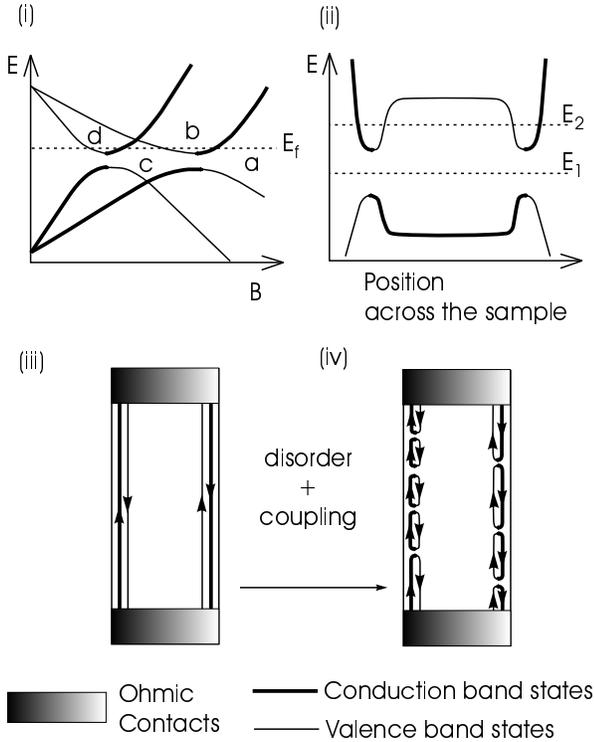}
\caption[figure 3]{(i) and (ii) the electron and hole Landau levels as a
function of magnetic field and spatial position respectively. \ Electron-like levels are
shown in heavier lines. (iii) and (iv) show possible edge current paths in
the insulating state with and without the presence of disorder.}
\end{figure}

In single carrier systems the quantum Hall condition also corresponds to a
situation where the Fermi level lies within a mobility gap between Landau
levels. As the levels approach the sample edges, however, they move upwards
in energy and form non-dissipative conducting edge states which generate
zeroes in the resistivity and quantum Hall plateaux. The picture can be
quite different in interacting electron-hole systems. Considering a spatial
plot of the energy levels of the electrons and holes shown in fig. 3ii we
expect that as they approach the edges of the structure the electron levels
will move upwards due to the edge confinement, while in contrast the hole
levels will move downwards. As a result the electron and hole levels will
always approach each other but due to the interband coupling the levels will
anticross, producing an energy gap. When the Fermi level in the bulk lies
between an equal number of electron and hole levels the structure will
always show completely insulating behaviour. This conclusion is obvious when
the Fermi level in the bulk also corresponds to the anticrossing gap at the
edges, e.g. for $E_{F}=E_{1}$ in Fig 3ii. It also holds when it crosses the
interacting electron and hole edge states since the number of these states
is equal leading to zero net current flow as for $E_{F}=E_{2}$ in Fig 3ii.
At contacts to the structure we expect edge-states to be at the potential of
the originating contact. Although the electron and hole states represent one
dimensional current paths with opposite directions (fig. 3iii), we might
expect that the edges would be equipotentials giving zero resistivity. This
is not the case however, since fluctuations in the potential could cause the
gap to rise above the Fermi level, thus disconnecting the channels into two
U-turns. More than one such disconnection will lead to a series of isolated 
conducting loops along the
sample edge. One possible arrangement is shown schematically in fig. 3iv. 

By contrast when there is population within the bulk of an unequal number of
Landau levels there is population of a finite net number of edge states. This gives
metallic behaviour with compensated quantum Hall plateaux in the Hall
resistance and zeros in $\rho _{xx}$. There is a rotation through $\thicksim $ 90$%
^{0}$ of the equipotentials which lie along the\ conducting edge
states of the bar in the metallic quantum Hall state but were mainly across the
bar for the insulating state where no Hall field was generated.

\begin{figure}[tbp]
\epsfxsize=3.4in
\epsffile{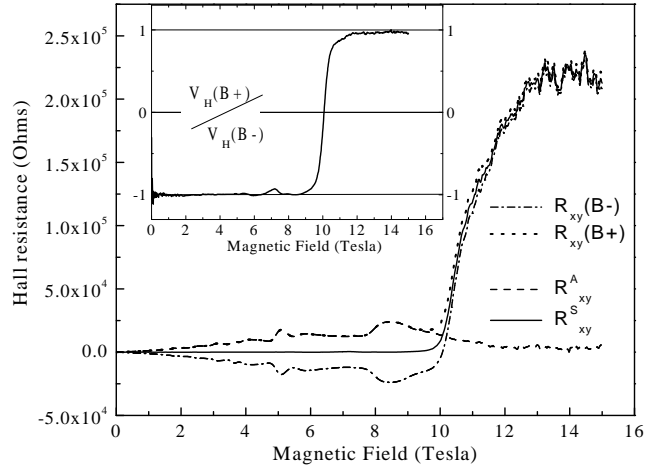}
\caption[figure 4]{The experimentally measured Hall resistance components $%
R_{xy}(B+)$ and $R_{xy}(B-)$, and the values deduced for the symmetric and
asymmetric Hall resistance $R_{xy}^{S}$ and $R_{xy}^{A}$ measured at 100mK.
The inset shows the ratio of the experimentally measured Hall resistivities
on reversing magnetic field direction.}
\end{figure}

The oscillations between insulating and metallic behaviour occur
as the Fermi level moves through regions b,d (metallic) and a,c (insulating)
in fig. 3.i. The appearance of an insulating state is crucially linked to the
anticrossing levels and it is interesting to note that a previous report of a re-entrant
insulating phase in p-SiGe was tentatively linked to unusual Landau level degeneracy\cite{Dunf}
where states may be crossing.

Evidence of a strong distortion of the current paths in the insulating
state comes from the large symmetric component (about 50\% of $%
\rho _{xx}$) in $R_{xy}$ occurring specifically at the fields where the system
becomes insulating. Fig. 4 shows $R_{xy}$ as measured with both directions
of the magnetic field. The symmetric and antisymmetric parts are also shown.
The inset shows the ratio $R_{xy}$($B+$)/$R_{xy}$($B-$). Up to $\thicksim $%
9T $R_{xy}$ is completely antisymmetric and the ratio is -1. The Hall
contacts show almost no admixing ($\leq $0.5\%) of the diagonal resistivity $%
\rho _{xx}$. At the onset of the insulating state however $R_{xy}$ switches
behaviour to become almost totally symmetric: the antisymmetric Hall
resistance tends to zero and a large finite voltage is developed across the
sample which does not depend on the polarity of the applied magnetic field
thus giving a large value of $R_{xy}^{S}$. This resistance has no obvious
functional relation to the diagonal resistivity $R_{xx}^{S}$ which rules out
a simple admixing origin. The Hall potential probes, although physically
opposite each other to an accuracy of within 10$\mu$m, are not
equipotentials despite the fact that $R_{xy}^{A}$=0.

Another striking feature of the symmetric part of $R_{xy}$ is the
reproducible resistance fluctuations which are accurately repeated for $B+$
and $B-$. These are shown in the inset to fig. 5, defined by $\Delta
R_{xy}^{S}=$ $R_{xy}^{S}-\overline{R}_{xy}^{S}$, where $\overline{R}%
_{xy}^{S} $ is the smoothed background. The functional form of the
resistance fluctuations remains essentially independent of temperature, only
increasing in magnitude. By contrast the Hall resistivity (=$R_{xy}^{A})$
shows much smaller fluctuations suggesting that any such disordered current
paths are almost completely symmetric with respect to reversal of the
direction of the carrier orbit. The violation of the Onsager relations
implies the non-local nature of the symmetric Hall voltage.

\begin{figure}[tbp]
\epsfxsize=3.15in
\epsffile{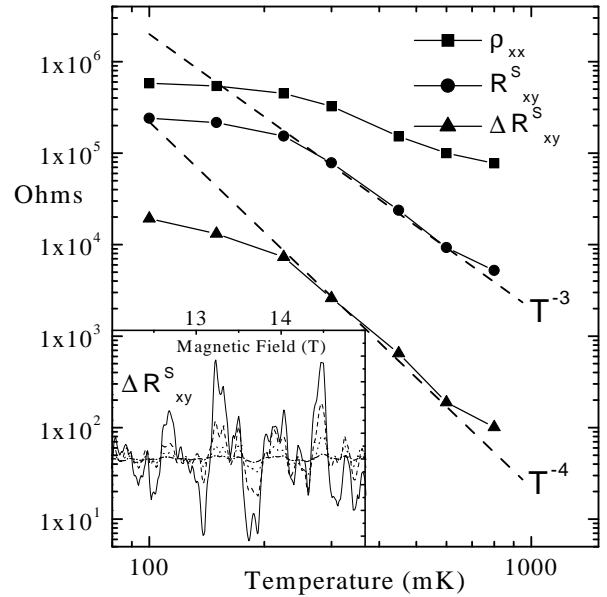}
\caption[figure 5]{The temperature dependence of $\rho _{xx}$, $%
R_{xy}^{S}$ and $\Delta R_{xy}^{S}$, showing power law behaviour for the symmetric components of the
Hall resistance. \ The inset shows $\Delta R_{xy}^{S}$ at 100, 225, 300 and 450 mK,
demonstrating that the form of the fluctuations is temperature independent.}
\end{figure}

The temperature dependence of the resistance components $\rho _{xx}$, $%
R_{xy}^{S}$ and $\Delta R_{xy}^{S}$ are shown in fig.5 for the insulating
region. The essentially non-Onsager form of the symmetric $R_{xy}$ component
leads us to analyse the directly measured resistance rather than
resistivity. The diagonal resistivity varies relatively slowly and saturates
below $\thicksim $100mK. The $R_{xy}^{S}$ and its fluctuations are more
strongly temperature dependent. Fits to these with exponentially
activated conduction only work over a factor of about two in
temperature range, however simple power laws work from the region of 800 to
200 mK where $R_{xy}^{S}$ is proportional to T$^{-3}$ and $\Delta R_{xy}$
to T$^{-4}$.

In conclusion, we have demonstrated that the electron-hole system oscillates
between insulating and conducting states as the magnetic field is increased.
A qualitative explanation has been proposed based on the formation of an
energy gap due to the anticrossing between the electron and hole
edge-states. The insulating states have been shown to have unusual behaviour
when the Hall resistance becomes symmetric with respect to field reversal.

{\bf Acknowledgements:} \ We are grateful to the UK-EPSRC and the EU for
continued support of this work.


\begin{references}
\bibitem{Klitzing}  K. von Klitzing, G. Dorda and M.Pepper, Phys. Rev. Lett. 
{\bf 45}, 494 (1980).

\bibitem{buttiker}  M. Buttiker, Phys. Rev. B. {\bf 38}, 9375 (1988)

\bibitem{chlovskii}  D.B. Chklovskii, B.I. Shklovskii, L.I. Glazman, Phys.
Rev. B. {\bf 46} 4026 (1992)

\bibitem{Naveh}  Y. Naveh and B. Laikhtman, Phys. Rev. Lett. {\bf 77}, 900
(1996)

\bibitem{Littlewood}  P.B. Littlewood and X. Zhu, Phys. Scr. {\bf T68}, 56
(1996)

\bibitem{deLeon}  S. de-Leon, L. D. Shvartsman, and B. Laikhtman, Phys. Rev.
B {\bf 60}, 1861 (1999)

\bibitem{Mendez}  E.E. Mendez et al, Phys. Rev. Lett. {\bf 55, }2216 (1985)

\bibitem{Daly}  M.S. Daly et al, Phys. Rev. B {\bf 53}, R10524 (1996).

\bibitem{symons}  D. M. Symons et al, Phys. Rev. B {\bf 58}, 7292 (1998).

\bibitem{Lakrimi}  M. Lakrimi et al, Phys. Rev. Lett. {\bf 152}, 3034 (1997).

\bibitem{Yang}  M. J. Yang, C.H. Yang, B.R. Bennett and B.V. Shanabrook,
Phys. Rev. Lett. {\bf 78}, 4613 (1997).

\bibitem{Dunf}  R.B Dunford et al, J. Phys.: Condens. Matter . {\bf 9}, 1565 (1997)

\bibitem{Nicholas}  R.J. Nicholas et al, Physica B {\bf 201}, 271 (1994).

\bibitem{Dalton}  K.S.H. Dalton et al, Surf. Sci. {\bf 156}, 305 (1994).

\bibitem{Nich2}  R.J. Nicholas et al, Physica E, in press (1999)

\bibitem{Altarelli}  M. Altarelli, Phys. Rev. B. {\bf 28}, 842 (1983).

\bibitem{Chiang}  Jin-Chen Chiang et al, Phys. Rev. Lett. {\bf 77}, 2053
(1997).

\bibitem{Cooper}  L.J. Cooper et al, Phys. Rev. B {\bf 57}, 11915 (1998)

\bibitem{Vasilev}  S.V. Vasil'ev et al, JEPS Lett. {\bf 69}, 343 (1999)

\bibitem{Petch}  C. Petchsingh et al, Physica E, (in press) (1999)


\end{references}
\end{document}